\documentclass[preprint,eqsecnum,prd,floatfix,onecolumn,preprintnumbers,oneside]{revtex4}
\usepackage{epsf}
\usepackage{epsfig}

\def\to{\rightarrow}

\def\ie{{\sl i.e.}}

\def\vb#1{\vbox to #1 pt{}}
\def\etal{{\it et.al.}}

\def\dofig#1#2{\epsfxsize=#1\centerline{\epsfbox{#2}}}

\def \lsim{\mathrel{\vcenter
     {\hbox{$<$}\nointerlineskip\hbox{$\sim$}}}}

\begin{document}
\preprint{\vbox{\baselineskip=14pt%
   \rightline{hep-ph/0409043}
}}

\title{NEUTRINOS IN ANOMALY MEDIATED SUPERSYMMETRY BREAKING WITH $R$--PARITY
  VIOLATION}

\author{F.~de Campos${}^a$, M.A.\ D\'{\i}az${}^b$, O.J.P.\ \'Eboli${}^c$, 
R.A.\ Lineros${}^b$, M.B.\ Magro${}^{c,d}$, \\ and P.G.\ Mercadante${}^c$.}
\affiliation{
${}^a$Departamento de F\'{\i}sica e Qu\'{\i}mica, Universidade Estadual 
Paulista, Av.\ Dr.\ Ariberto Pereira da Cunha 333, Guaratinguet\'a, SP, 
Brazil
}

\affiliation{
${}^b$Departamento de F\'{\i}sica, Universidad Cat\'olica de Chile,
Av. V. Mackenna 4860, Santiago, Chile
}

\affiliation{
${}^c$Instituto de F\'{\i}sica da USP, C.P.\ 66.318, 
05315--970, S\~ao Paulo, Brazil
}

\affiliation{
${}^d$Faculdade de Engenharia, Centro Universit\'ario FSA,
Av.\ Pr\'{\i}ncipe de Gales 821, Santo Andr\'e, SP, Brazil
}

%


\begin{abstract}
  
  We show that a supersymmetric standard model exhibiting anomaly
  mediated supersymmetry breaking can generate naturally the observed
  neutrino mass spectrum as well mixings when we include bilinear
  $R$--parity violation interactions.  In this model, one of the
  neutrinos gets its mass due to the tree level mixing with the
  neutralinos induced by the $R$--parity violating interactions while
  the other two neutrinos acquire their masses due to radiative
  corrections. One interesting feature of this scenario is that the
  lightest supersymmetric particle is unstable and its decay can be
  observed at high energy colliders, providing a falsifiable test of
  the model.

\end{abstract}

\maketitle

\bigskip



\section{Introduction}

There have been many experimental results in neutrino physics \cite{neutrinos}
which have established the pattern of neutrino oscillation and masses, clearly
requiring physics beyond the standard model (SM) to explain it. In general,
neutrino oscillations are parametrized by three mixing angles, with two of
them large, as opposed to the quark sector where the mixing angles are all
small.  Moreover, neutrino oscillations depend also on the mass squared
differences, and we have learned that the mass difference corresponding to the
atmospheric neutrino oscillations is much larger than the corresponding one to
the solar neutrino oscillations \cite{neutrinos}.  The analysis of the solar
neutrinos leads to the $3\sigma$ level limits \cite{neut:pheno} (see also
\cite{more-nu})
\begin{eqnarray}
0.23 \lsim & \sin^2\theta_{\mathrm{sol}} & \lsim 0.37 \cr
7.3\times10^{-5} \lsim & \Delta m^2_{\mathrm{sol}} & \lsim 9.1\times10^{-5}
\,{\mathrm{eV}}^2
\label{neut:solar}
\end{eqnarray}
while the atmospheric neutrino data shows that
\begin{eqnarray}
0.34 \lsim & \sin^2\theta_{\mathrm{atm}} & \lsim 0.66 \cr
1.4\times10^{-3} \lsim & \Delta m^2_{\mathrm{atm}} & \lsim 3.3\times10^{-3}
\,{\mathrm{eV}}^2  \; .
\label{neut:atm}
\end{eqnarray}

Despite the lack of direct experimental evidence on supersymmetry (SUSY),
supersymmetric models are promising candidates for physics beyond the SM. It
is an experimental fact that SUSY must be broken if it is realized in nature.
Here, we consider the anomaly--mediated SUSY breaking scenario (AMSB) where
the supersymmetry breaking in the hidden sector is transmitted into the
observable sector by the Super-Weyl anomaly \cite{amsb}. Below the
compactification scale, the model is described by an effective four
dimensional supergravity theory, where the soft supersymmetry breaking
parameters are generated beyond tree level.  Without further contributions the
slepton squared masses turn out to be negative. This tachyonic slepton problem
can be solved by adding a common scalar mass to all scalars
\cite{Kaplan:2000jz}.

It has been suggested a long time ago that supersymmetry and neutrino masses
and mixings may be deeply tied together \cite{neut:rp}.  One way of giving
mass to the neutrinos in supersymmetric models is via Bilinear $R$--Parity
Violation (BRpV): in such model, bilinear terms which violates lepton number
as well as $R$--Parity are introduced in the superpotential 
\cite{brpv:orig,Akeroyd:1997iq,Carvalho:2002bq}.
As a consequence, one neutrino acquires mass at tree level due to a low energy
see--saw mechanism in which neutrinos mix with neutralinos. The other two
neutrinos become massive via one--loop corrections to the neutralino--neutrino
mass matrix \cite{brpv:oneloop}.

It has been shown that BRpV can be successfully embedded into models with AMSB
\cite{DeCampos:2001wq} giving rise at tree level to a neutrino mass compatible
with the atmospheric neutrino mass scale.  Here, we generalize this model
including lepton number violating interaction in the three generations and we
show that the inclusion of one--loop corrections to the neutralino--neutrino
mass matrix leads to a neutrino spectrum and mixings compatible with the
available data.  This is non--trivial since the radiative corrections depend
on the SUSY spectrum and it is not clear {\em a priori} that the corrections
will have the required properties after we impose the existing limits on the
SUSY mass spectrum.

This paper is organized as follows. In Section \ref{brpvmodel} we briefly show
the principles of the AMSB-BRpV model. In Section \ref{masses} we discuss the
effects of BRpV on the neutrino mass and mixing angles. In Section \ref{punto}
we present our reference scenario and its main properties.  The
results for an AMSB model exhibiting BRpV are presented in Section
\ref{results} and we conclude in Section \ref{bye}.

\section{The AMSB--BRpV model}
\label{brpvmodel}

The superpotential of our BRpV model includes three $\epsilon_i$ parameters
with units of mass not present in the MSSM \cite{brpv:orig,Diaz:1997xc}
\begin{equation}
     W=W_Y-\mu\widehat H_d\widehat H_u+\epsilon_i\widehat L_i\widehat H_u
\; ,
\end{equation}
with $W_Y$ including the Yukawa interactions. The $\epsilon$ terms violate
lepton number and $R$--Parity and satisfy $|\epsilon_i| \ll \mu$.  The
appearance of $R$--Parity violating bilinear terms and not trilinear terms can
be justified in models with a horizontal symmetry \cite{Mira:2000gg}, or in
models with spontaneous $R$--Parity breaking \cite{Romao:1996mg}. In addition
to the $\epsilon_i$ terms in the superpotential, we must add the soft
lagrangian bilinear terms proportional to $B_i\epsilon_i$.  For our purposes
the relevant terms are
\begin{equation}
  V_{soft}\owns -B\mu H_dH_u+B_i\epsilon_iL_iH_u+
  m_{H_d}^2 H^*_dH_d+m_{L_i}^2\widetilde L^*_i\widetilde L_i  \; .
\end{equation}
In principle, it is possible to redefine the superfields in order to make BRpV
terms disappear from the superpotential, however, they do not vanish from the
soft lagrangian simultaneously \cite{Diaz:1997vq}.

The scalar potential of BRpV models is such that the sneutrino fields acquire
a non zero vacuum expectation value $v_i$ that leads to the generation of
neutrino mass and mixing angles.  Even in the basis where the $\epsilon_i$
terms are removed from the superpotential, the sneutrino fields acquire
non-zero vev's $v^\prime_i\approx(\epsilon_i v_d+\mu v_i)/\mu$
\cite{brpv:oneloop} which, from the minimization of the scalar potential, can
be shown to satisfy
\begin{equation}
    v^\prime_i\approx 2 v_d\,{{\epsilon_i}\over{\mu}}\,\frac
    {\Delta m_i^2 - \mu \Delta B_i \tan\beta }{2 M^2_{L_i}+m_Z^2 \cos(2\beta)}
\end{equation}
with $\Delta m^2_i=M^2_{L_i}-m_{H_d}^2$ and $\Delta B_i=B_i-B$. Since there is
no reason to assume that these differences are zero {\em at the weak scale},
the sneutrino vev's are not zero either.

The parameters defining our BRpV--AMSB model are the usual ones in AMSB models
\begin{equation}
  m_0          \;\; , \;\;
  m_{3/2}      \;\; , \;\;
  \tan\beta  \;\; , \hbox{ and } \;\;
  \hbox{sign}(\mu)  \; ,
\end{equation}
where the scalar mass $m_0$ and the gravitino mass $m_{3/2}$ are given at the
unification scale. This set of parameters is supplemented by the six BRpV
parameters $\epsilon_i$ and $B_i$. It is advantageous to trade $B_i$ by a
parameter more directly connected to the neutrino physics observables,
therefore, we shall choose the parameters $\Lambda_i$ defined below instead of
$B_i$ as input parameters. One of the virtues of AMSB models is that the
$SU(2) \otimes U(1)$ symmetry is broken radiatively by the running of the
parameters from the GUT scale to the weak one. This feature is preserved in
our model exhibiting BRpV.

\section{Neutrino and Neutralino Masses}
\label{masses}

In the basis $\psi^{0T}= (-i\lambda^\prime, -i\lambda^3, \widetilde{H}_d^1,
\widetilde{H}_u^2, \nu_{e}, \nu_{\mu}, \nu_{\tau} )$ the $7 \times 7$ neutral
fermion mass matrix $M_N$ has the see--saw structure, at tree level,
\begin{equation}
M_N=\left[  
\begin{array}{cc}  
{\cal M}_{\chi^0}& m^T \cr
\vb{20}
m & 0 \cr
\end{array}
\right] \; ,
\end{equation}
where the standard MSSM neutralino mass matrix is
\begin{equation}
{\cal M}_{\chi^0}\hskip -2pt=\hskip -4pt \left[ \hskip -7pt 
\begin{array}{cccc}  
M_1 & 0 & -\frac 12g^{\prime }v_d & \frac 12g^{\prime }v_u \cr
\vb{12}   
0 & M_2 & \frac 12gv_d & -\frac 12gv_u \cr
\vb{12}   
-\frac 12g^{\prime }v_d & \frac 12gv_d & 0 & -\mu  \cr
\vb{12}
\frac 12g^{\prime }v_u & -\frac 12gv_u & -\mu & 0  \cr
\end{array}  
\hskip -6pt
\right]  \; , 
\end{equation}
and $R$--parity violating interactions give rise to sneutrino vev's and
mixings between neutrinos and gauginos/higgsinos:
\begin{equation}
m=\left[  
\begin{array}{cccc}  
-\frac 12g^{\prime }v_1 & \frac 12gv_1 & 0 & \epsilon _1 \cr
\vb{18}
-\frac 12g^{\prime }v_2 & \frac 12gv_2 & 0 & \epsilon _2  \cr
\vb{18}
-\frac 12g^{\prime }v_3 & \frac 12gv_3 & 0 & \epsilon _3  \cr  
\end{array}  
\right] \; .
\end{equation}
Due to its structure $M_N$ exhibits just one massive neutrino at tree level
while the other two remain massless. The degeneracy of these two states is
lifted when we include the one--loop corrections to the neutral fermion mass
matrix.

In order to get some intuition on the main effects of BRpV, it is instructive
to analyze the limit $|\epsilon_i| \ll \mu$ where we can perform a
perturbative diagonalization of the neutralino--neutrino mass matrix in terms
of the parameters \cite{brpv:orig}
\begin{equation}
  \xi \equiv m \cdot {\cal M}^{-1}_{\chi^0} \; .
\end{equation}
In this approximation $M_N$ is diagonalized by the rotation matrix
\begin{equation}
  {\cal N}^* \simeq \left (
\begin{array}{rc}
N^* & N^* \xi^\dagger \\
- V^T_\nu \xi & V^T_\nu
\end{array}
\right ) \; ,
\end{equation}
where $N^*$ diagonalizes the $4 \times 4$ neutralino mass matrix ${\cal
  M}_{\chi^0}$ and $V_\nu$ diagonalizes the effective tree level neutrino mass
matrix
\begin{equation}
  {\bf M}^{\rm eff} = - m ~{\cal M}^{-1}_{\chi^0}~ m^T \; .
\end{equation}

Within this approximation \cite{brpv:oneloop}, it can be shown that the
atmospheric mass scale is adequately described by the tree level neutrino mass
\begin{equation}
m_{\nu_3}^{\rm tree}={{M_1g^2+M_2g'^2}\over{4\Delta_0}}|\vec\Lambda|^2 \; ,
\label{nutree}
\end{equation}
where $\Delta_0$ is the determinant of the neutralino sub--matrix ${\cal
  M}_{\chi^0}$ and $\vec\Lambda=(\Lambda_1,\Lambda_2,\Lambda_3)$, with
\begin{equation}
\Lambda_i=\mu v_i+\epsilon_i v_d \approx \mu v^\prime_i \; ,
\label{lambda}
\end{equation}
and the index $i$ refers to the lepton family. Due to this direct relation
between neutrino mass and $\Lambda_i$, from now on, we eliminate the $B_i$ as
independent parameters in favor of $\Lambda_i$.  Moreover, the mixing angles
between the massive neutrino state and the tree level massless ones are given
approximately by
\begin{equation}
   \tan \theta_{13} = - \frac{\Lambda_1}{\sqrt{\Lambda_2^2 + \Lambda_3^2}} 
  \;\;\;\; \hbox{ and } \;\;\;\;
   \tan \theta_{23} = - \frac{\Lambda_2}{\Lambda_3} \; .
\label{mix:atm}
\end{equation}

The one--loop corrections to the neutralino--neutrino mass matrix in the
presence of BRpV interactions have been evaluated in Refs.\ 
\cite{brpv:oneloop}. The mixing of neutrinos with gauginos and higgsinos due
to BRpV leads to effective interactions neutrino--quark--squark and
neutrino--lepton--slepton.  These vertices give rise to squark--quark and
slepton--lepton loop contributions to the neutrino mass matrix that are
proportional to the $R$--Parity violating parameters, vanishing when
$R$--Parity is conserved \cite{brpv:oneloop}.  In the approximation that only
the bottom quark--bottom squark loop gives a sizeable contribution to the
one--loop corrections to the neutrino masses \cite{Diaz:2003as}, the scale of
the solar neutrino mass is given by
\begin{equation}
   m_{\nu_2} \propto \frac{ | \vec{\epsilon} |^2}{16 \pi^2 \mu^2} m_b
   \; ,
\end{equation}
while the solar mixing angle is
\begin{equation}
   \tan \theta_{12} \simeq \left |  \frac{\tilde{\epsilon}_1}
     {\tilde{\epsilon}_2}     \right |
    \;\;\;\; \hbox{ where } \tilde{\epsilon_i} = V^{\nu,~{\rm tree}}_{ij}
      \epsilon_j \; .
\label{mix:solar}
\end{equation}
Still within this approximation we have
\begin{eqnarray}
\label{eq:TildeEpsilon}
 \tilde\epsilon_1&=&{{
 \epsilon_1(\Lambda_2^2+\Lambda_3^2)-\Lambda_1
 (\Lambda_2\epsilon_2+\Lambda_3\epsilon_3)
  }\over{
\sqrt{\Lambda_2^2+\Lambda_3^2}~~
\sqrt{\Lambda_1^2+\Lambda_2^2+\Lambda_3^2}
}} \; ,
\nonumber\\
\tilde\epsilon_2&=&{{
\Lambda_3\epsilon_2-\Lambda_2\epsilon_3
}\over{
\sqrt{\Lambda_2^2+\Lambda_3^2}
}} \; ,
\\
\tilde\epsilon_3&=&{{
\vec\Lambda\cdot\vec\epsilon
}\over{
\sqrt{\Lambda_1^2+\Lambda_2^2+\Lambda_3^2}
}} \; . \nonumber
\end{eqnarray}
These approximations are valid when the tree level contribution is dominant
and one--loop contributions are small corrections. However, this is not always
true as we will see in the next sections. In this case, the above formulas are
no longer valid.

\section{Reference Scenario}
\label{punto}

In order to understand the main features of our model, we considered a point
in the parameter space which satisfies all the collider and neutrino physics
constraints and then explore the parameter space around it.  First, we choose
an AMSB scenario in which all superpartner masses satisfy the present
experimental bounds and where we obtain a correct electroweak symmetry
breaking; we used the code SUSPECT \cite{Djouadi:2002ze} to compute the
two--loop running of the parameters from the GUT scale to the weak one. At
this scale, the minimization of the scalar potential is generalized to include
five vacuum expectation values: $v_u$ and $v_d$ for the Higgs fields and $v_i$
for the sneutrinos. For the AMSB parameters, see experimental restrictions on
the parameters in \cite{Abdallah:2003gv}, we chose
\begin{equation}
  m_{3/2}=35 \,{\mathrm{TeV}}\,,\quad 
  m_0=250 \,{\mathrm{GeV}}\,,\quad 
  \tan\beta=15\,,\quad \hbox{ and} \quad
 {\mathrm{sign}}(\mu) < 0 \; .
\label{refAMSB}
\end{equation}
Second, we randomly varied the parameters $\epsilon_i$ and $\Lambda_i$ looking
for solutions in which the restrictions (\ref{neut:solar}) and
(\ref{neut:atm}) from neutrino physics are satisfied. An example of these
solutions is
\begin{eqnarray}
& \epsilon_1 = -0.015 \hbox{ GeV}\;,  \qquad & \Lambda_1 = -0.03  \hbox{
  GeV}^2 \; ,
\nonumber\\
& \epsilon_2 = -0.018 \hbox{ GeV}\;,   \qquad & \Lambda_2 = -0.09  \hbox{
  GeV}^2 \; , 
\label{basic:brpv}
\\
& \epsilon_3 = 0.011 \hbox{ GeV}\;,  \qquad & \Lambda_3 = -0.09   \hbox{
  GeV}^2  \; . \nonumber
\end{eqnarray}
Eqs.~(\ref{refAMSB}) and (\ref{basic:brpv}) define what we call our reference
model. The neutrino parameters obtained in this reference model are
\begin{eqnarray}
 \Delta m^2_{\mathrm{atm}}=2.4\times10^{-3}\,{\mathrm{eV}}^2\,,\qquad & &
 \tan^2\theta_{\mathrm{atm}}=0.72\;, 
\nonumber
\\
 \Delta m^2_{\mathrm{sol}}=7.9\times10^{-5}\,{\mathrm{eV}}^2\,,\qquad & &
 \tan^2\theta_{\mathrm{sol}}=0.47\;, 
\label{NeuRes}
\\
 & & \tan^2 \theta_{13} = 0.033 \; ,
\nonumber
\end{eqnarray}
which agree with the present experimental results. 

The effective neutrino mass matrix in our reference model has the structure
\begin{equation}
  {\bf M}^{\rm eff}=\left[\matrix{
  \lambda  & 2\lambda & \lambda \cr
  2\lambda & a        & b       \cr
  \lambda  & b        & m
   }\right] \; ,
\label{texture}
\end{equation}
where $m \approx 0.031$ eV sets the overall scale, $a/m\approx 0.74$, $b/m
\approx 0.67$, and $\lambda/ m \approx 0.12-0.14$. As a first approximation,
we can neglect the effect of $\lambda$ and estimate the two heavier neutrino
masses
\begin{equation}
   m_{\nu_{2,3}}=\frac{1}{2}\left[m+a\pm\sqrt{(m-a)^2+4b^2}\right]
\label{NuMassApp}
\end{equation}
with the lightest neutrino being nearly massless. Numerically, the
approximation in Eq.~(\ref{NuMassApp}) leads to $\Delta m^2_{\mathrm{atm}}
\approx 2.3 \times10^{-3}{\mathrm{eV}}^2$ which is very close to the complete
result given in Eq.~(\ref{NeuRes}). Still within this approximation, the
atmospheric mixing angle satisfy
\begin{equation}
   \tan2\theta_{\mathrm{atm}}\approx\frac{2b}{m-a}
\end{equation}
which leads $\tan^2\theta_{\mathrm{atm}} \approx 0.68$, also in close
agreement with the complete result shown in Eq.~(\ref{NeuRes}).  However, for
the solar mass difference, the $\lambda=0$ approximation is not good giving
near half the correct value in Eq.~(\ref{NeuRes}).

In general, the effective neutrino mass matrix at one--loop has the
approximated form \cite{Diaz:2003as}
\begin{equation}
 {\bf M}^{\rm eff}_{ij} = A\Lambda_i\Lambda_j+
  B(\epsilon_i\Lambda_j+\epsilon_j\Lambda_i)+C\epsilon_i\epsilon_j \; ,
\label{deltapi}
\end{equation}
where the coefficient $A$ receives tree--level as well as one--loop
contributions, and $B$ and $C$ are one--loop generated. Approximated
expressions for the contributions from bottom/sbottom and
charged--scalar/charged--fermion loops to the parameters $A$, $B$ and $C$ can
be found in \cite{Diaz:2003as}. For our reference model, we have $A \approx 3
\,\,{\mathrm{eV}} / {\mathrm{GeV}}^4$, $B \approx-2\,\, {\mathrm{eV}} /
{\mathrm{GeV}}^3$, and $C\approx 15 \,\, {\mathrm{eV}} /{\mathrm{GeV}}^2$.
Therefore, for our reference point, the one--loop generated parameters $B$ and
$C$ are not negligible, stressing the necessity of using one--loop corrected
expressions.

Some of the entries in the effective neutrino mass (\ref{deltapi}) are
numerically small for our reference model, which allow us to write as a good
first approximation that
\begin{equation}
 {\bf M}^{\mathrm{eff}}=\left[\matrix{
 A\Lambda_1^2+2B\Lambda_1\epsilon_1+C\epsilon_1^2 & \cdot & \cdot \cr
 A\Lambda_1\Lambda_2+B(\epsilon_1\Lambda_2+\epsilon_2\Lambda_1)
 +C\epsilon_1\epsilon_2 & A\Lambda_2^2+2B\epsilon_2\Lambda_2  & \cdot  \cr
 A\Lambda_1\Lambda_3+C\epsilon_1\epsilon_3 & A\Lambda_2\Lambda_3 & 
 A\Lambda_3^2+2B\epsilon_3\Lambda_3
 }\right] \; ,
\label{massmat}
\end{equation}
where the matrix is symmetrical and we do not repeat the redundant terms. From
this expression, it can be shown, neglecting the first row and the first
column, that the atmospheric angle is given by
\begin{equation}
  \tan2\theta_{\mathrm{atm}}=\frac{2A\Lambda_2\Lambda_3}
  {A(\Lambda_3^2-\Lambda_2^2)+2B(\epsilon_3\Lambda_3-\epsilon_2\Lambda_2)}
  \; .
\label{tanatmapp}
\end{equation}
Since only $A$ receives contributions at tree--level, it is no surprise that
the above formula approaches the tree--level expression in Eq.~(\ref{mix:atm})
when $B\rightarrow0$. Nevertheless, as we have seen, the one--loop corrections
to $B$ are non-negligible in our reference model, leading to departures from
the na\"\i ve expectations.  Moreover, we can also demonstrate that the masses
of the two heaviest neutrinos are approximated by
\begin{equation}
 m_{\nu_{2,3}}=
 \textstyle{\frac{1}{2}}A(\Lambda_3^2+\Lambda_2^2)+
 B(\epsilon_3\Lambda_3+\epsilon_2\Lambda_2)
 \pm\sqrt{\left[
\textstyle{\frac{1}{2}}A(\Lambda_3^2-\Lambda_2^2)+
 B(\epsilon_3\Lambda_3-\epsilon_2\Lambda_2)\right]^2+
 A^2\Lambda_2^2\Lambda_3^2}\;\;,
\label{numasses}
\end{equation}
while the lightest neutrino mass is negligible.  Since the values of
$\epsilon_2$ and $\epsilon_3$ are much smaller than $\Lambda_2=\Lambda_3$ in
our reference model, the $A$ term gives rise to the most important
contribution. The two neutrino masses in Eq.~(\ref{numasses}) are
hierarchical, therefore the squares of $m_{\nu_3}$ and $m_{\nu_2}$ are a good
first approximation to the atmospheric and solar mass squared differences.

\section{Results}
\label{results}

\begin{figure}
\dofig{5in}{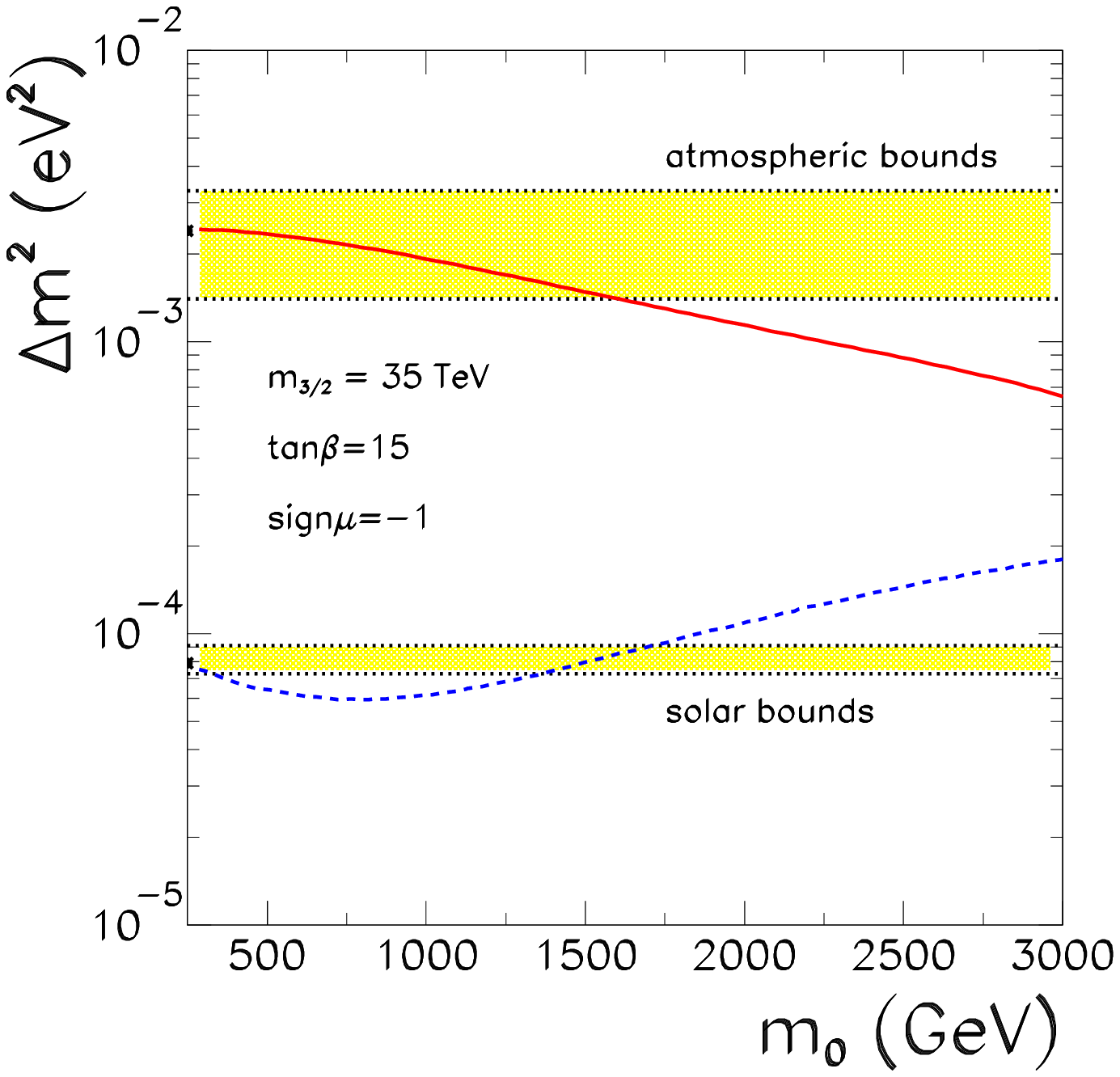}
\vskip .5cm
\caption[]{
  The red solid (blue hashed) line stands for the predicted atmospheric
  (solar) mass squared difference as a function of the scalar mass $m_0$ for
  $m_{3/2}=35$ TeV, $\tan\beta=15$, and $\hbox{sign}(\mu) < 0$ and for the
  BRpV parameters given in (\ref{basic:brpv}). The allowed $3\sigma$
  atmospheric (solar) mass squared difference is represented by the upper
  (lower) horizontal yellow band. Our reference point is represented by a star
  on the left of the plot.  }
\label{m0-1}
\end{figure}

Let us initially analyze the dependence of the neutrino masses and mixings
upon the AMSB parameters. In Fig.~\ref{m0-1} it is shown the dependence on the
scalar mass $m_0$ of the predicted atmospheric neutrino mass squared
difference $\Delta m^2_{\mathrm{atm}}$ (red solid line) and the solar neutrino
mass squared difference $\Delta m^2_{\mathrm{\mathrm{sol}}}$ (blue dashed
line), for fixed values $m_{3/2}=35$ TeV, $\tan\beta=15$, and $\hbox{sign}
(\mu) < 0$ and for the BRpV parameters given in (\ref{basic:brpv}).  As we can
see from this figure, $\Delta m^2_{\mathrm{atm}}$ is a decreasing function of
$m_0$ while $\Delta m^2_{\mathrm{\mathrm{sol}}}$ has a more complex behavior.

When the scalar mass $m_0$ increases the masses of the scalar SUSY particles
grow, and consequently, decoupling effects make the $A$ term to diminish. In
fact all three terms $m$, $a$, and $b$ in the texture (\ref{texture}) decrease
with $m_0$. This explains why the atmospheric mass decreases monotonically
with $m_0$ in Fig.~\ref{m0-1}. Nevertheless, this phenomenon does not happen
for the solar mass due to the minus sign in Eq.~(\ref{numasses}) which causes
the complex behavior. For instance, at large $m_0$ the square root decreases a
slightly faster than the term outside the square root, increasing the solar
mass as observed in Fig.~\ref{m0-1}.

We can also see from Fig.~\ref{m0-1} that $\Delta m^2_{\mathrm{atm}}$ is
within the present experimental bounds for $m_0 \lesssim 1.6$ TeV while
$\Delta m^2_{\mathrm{sol}}$ satisfies the experimental constraints for $m_0
\lesssim 310$ GeV and $ 1.4 \hbox{ TeV} \lesssim m_0 \lesssim 1.75$ TeV.
Therefore, our models lead to acceptable neutrino masses provided $m_0
\lesssim 310$ GeV or $ 1.4 \hbox{ TeV} \lesssim m_0 \lesssim 1.6$ TeV for all
other parameters fixed at their reference values.  It is also important to
notice that the heaviest neutrino state has a mass of the order of $0.050$ eV
for our reference point and that it decreases as $m_0$ increases. Moreover,
the radiative corrections lead to a contribution of ${\cal O}(10\%)$,
therefore, the tree--level result for the neutrino mass is a good order of
magnitude estimative.

\begin{figure}
\dofig{5in}{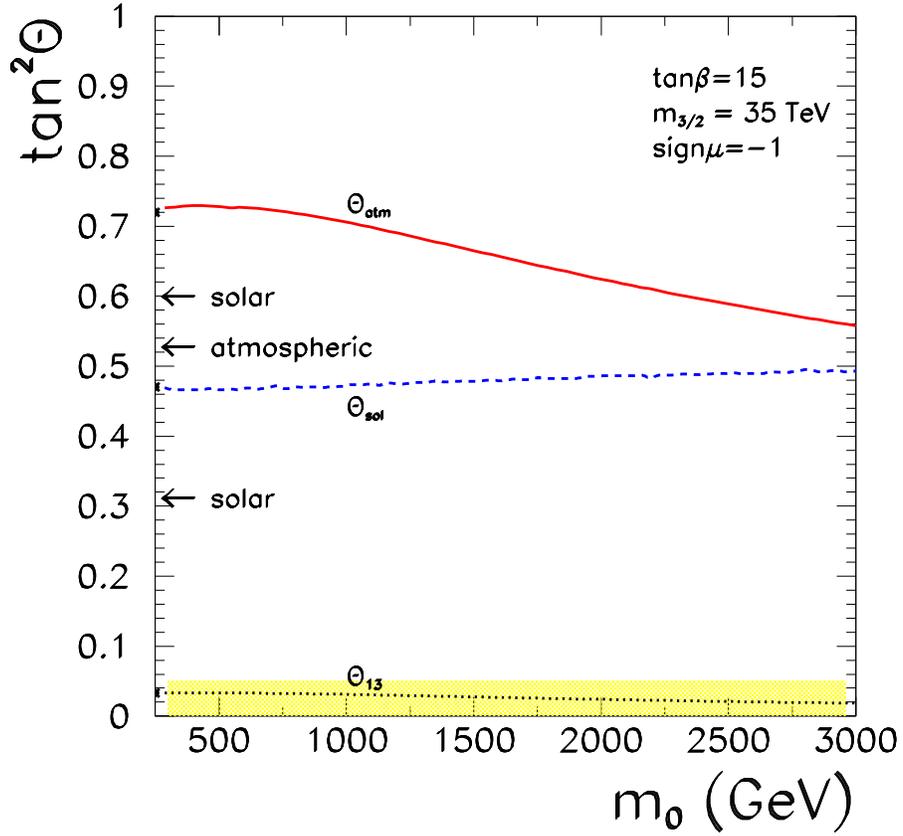}
\vskip .5cm
\caption[]{
  Predictions of the AMSB--BRpV model for the atmospheric, solar, and
  $\theta_{13}$ mixing angles as a function of the scalar mass $m_0$ for the
  same parameters as in Fig.~\ref{m0-1}. The upper and lower $3\sigma$ bounds
  on the solar mixing angle and the $3\sigma$ lower bound on the atmospheric
  mixing angle are marked by the arrows while the yellow horizontal band
  states for the $\theta_{13}$ allowed region. Our reference point is
  represented by a star on the left of the plot. }
\label{m0-3}
\end{figure}

We depict in Fig.~\ref{m0-3} the tangent squared of the atmospheric (solar)
angle $\tan^2\theta_{\mathrm{atm (sol)}}$ as a function of $m_0$ for our
reference point. As we can see from this figure, there is a small dependence
on $m_0$ of the atmospheric mixing angle and a milder one of the solar mixing.
Due to the importance of one--loop corrections, the lowest order approximated
expressions (\ref{mix:atm}) and (\ref{mix:solar}) do not describe the
dependence of the solar and atmospheric mixings on the scalar mass $m_0$; in
order to understand this behavior we should use the full one--loop
approximation (\ref{tanatmapp}). For $\Lambda_2=\Lambda_3$, this expression
reduces to
\begin{equation}
 \tan2\theta_{\mathrm{atm}}=\frac{A\Lambda_3}  {B(\epsilon_3-\epsilon_2)}\,.
\label{tanatmapp2}
\end{equation}

We checked that there is a clear decrease of the parameter $A$ with $m_0$,
while $B$ slightly increases, explaining the slope in the atmospheric angle
curve. We verified further that a similar effect happens for $m_{3/2}$, \ie, a
very mild dependence of the solar angle and a slight decrease of $\tan^2
\theta_{\mathrm{atm}}$ with increasing $m_{3/2}$, since its tree level
contribution to $A$ is inversely proportional to the gaugino masses.  Note
from Eqs.~(\ref{tanatmapp}) and (\ref{tanatmapp2}) that the fact that
$\epsilon_2$ and $\epsilon_3$ have different signs is responsible for
preventing the atmospheric angle to be maximal since $\Lambda_2=\Lambda_3$ in
our reference model.

\begin{figure}
\dofig{5in}{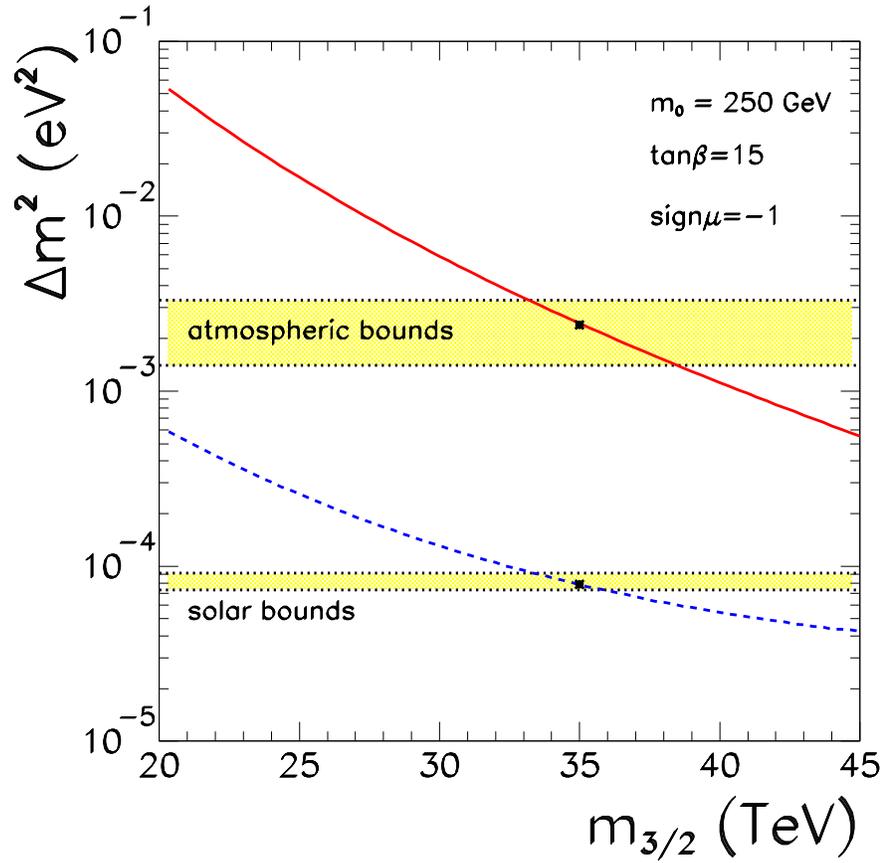}
\vskip .5cm
\caption[]{
  Atmospheric (solid line) and solar (dashed line) mass squared differences as
  a function of the gravitino mass $m_{3/2}$. The remaining parameters assume
  the value of our reference point and the conventions are the same of
  Fig.~\ref{m0-1}.  }
\label{m32-1}
\end{figure}

In Fig.~\ref{m32-1} we display the dependence of the atmospheric and solar
mass squared differences on the gravitino mass $m_{3/2}$ for the other
parameters assuming their reference values. First of all, the observed
dependence is much stronger compared to the dependence on $m_0$; this is
expected due to the large impact of $m_{3/2}$ on the soft gaugino masses,
which together with $\mu$ define the tree--level neutrino mass matrix.
Moreover, the SUSY spectrum has a large impact on the one--loop corrections
increasing the sensitivity to $m_{3/2}$.  Both solar and atmospheric squared
mass differences are too large in the region of small gravitino masses,
however, this region is already partially ruled out since it leads to
charginos lighter than the present experimental bounds for $m_{3/2} \lesssim
30$ TeV.  Conversely, there is no acceptable solution for the neutrino masses
at large $m_{3/2}$, again a region partially ruled out by data since the staus
are too light in this region.  Furthermore, we can see from this figure that
our AMSB--BRpV model leads to acceptable neutrino masses for a small window of
the gravitino mass ($33\; \rm{TeV}\lesssim m_{3/2} \lesssim 36\;\rm{TeV}$)
given our choice of parameters. This is far from trivial since we have no {\em
  a priori} guaranty that we can generate the required neutrino spectrum,
specially the radiative corrections, satisfying at the same time the
experimental constraints on the superpartner masses.

\begin{figure}
\dofig{5in}{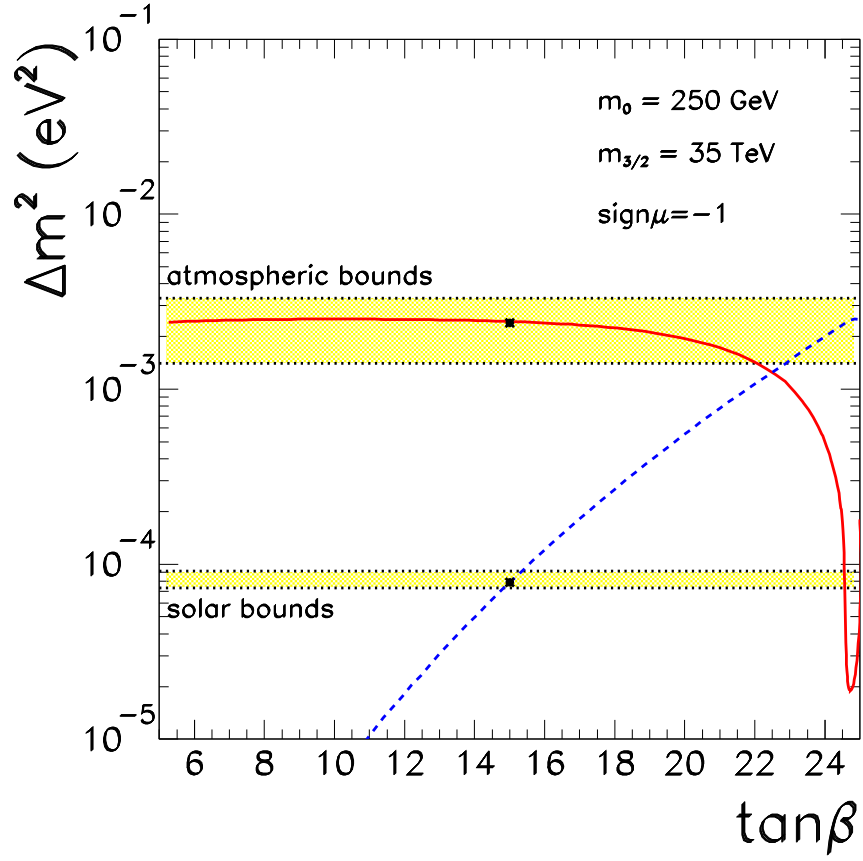}
\vskip .5cm
\caption[]{
  Atmospheric (solid line) and solar (dashed line) mass squared differences as
  a function of $\tan\beta$.  The remaining parameters assume the value of our
  reference point and the conventions are as in Fig.~\ref{m0-1}.  }
\label{tgb-1}
\end{figure}

In Fig.~\ref{tgb-1} we present $\Delta m^2_{\mathrm{atm}}$ and $\Delta
m^2_{\mathrm{sol}}$ as a function of $\tan \beta$ with all other parameters
fixed at their reference values. Clearly, $\Delta m^2_{\mathrm{atm}}$ has a
very mild dependence upon $\tan \beta$ since it is dominated by the tree level
contributions.  Nevertheless, $\Delta m^2_{\mathrm{sol}}$ presents a strong
dependence on this parameter, exhibiting an acceptable solution only for a
very narrow range $14.8 \lesssim \tan\beta \lesssim 15.3$. This is a
consequence of the strong dependence of the radiative corrections on
$\tan\beta$.  The effective interaction $\nu b \tilde{b}$ generated after the
diagonalization of the neutralino--neutrino mass matrix has a coupling
$\lambda^\prime_{333}$ that exhibits a term proportional to $\tan \beta$,
leading to the observed strong dependence. In other words, the neutrino mixes
with the higgsino, which couples to bottom-sbottom via the bottom Yukawa
coupling which increases with $\tan\beta$. At large $\tan\beta$, the lightest
stau become lighter than the present experimental bounds, therefore, this
region is already experimentally excluded.

\begin{figure}
\dofig{5in}{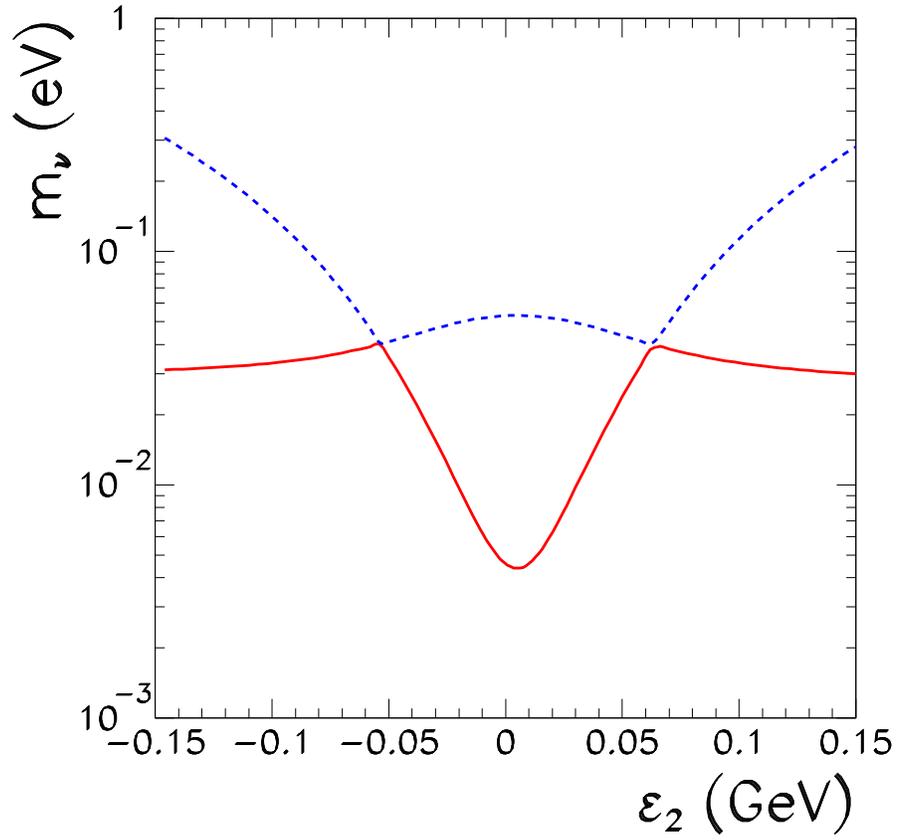}
\vskip .5cm
\caption[]{ Eigenvalues of the effective neutrino mass matrix as a
  function of $\epsilon_2$ with the other parameters fixed at the reference
  point. The smallest eigenvalue is always smaller than $5 \times 10^{-5}$ eV.
}
\label{eigen}
\end{figure}

The neutrino masses and mixings present a rich structure when we vary the BRpV
parameters. We display in Fig.~\ref{eigen} the neutrino masses as a function
of $\epsilon_2$ keeping all other parameters fixed at the reference value; the
smallest mass eigenvalue is not displayed and it is usually smaller than $5
\times 10^{-5}$ eV. As we can see, there is one eigenvalue of the effective
neutrino mass matrix that is approximately constant since it is associated to
the tree level neutrino mass. However, as $|\epsilon_2|$ grows the radiative
corrections start to become important and even to dominate; we observe the
same behavior of the neutrino masses with $\epsilon_1$ and $\epsilon_3$. This
is the origin of the level crossing that we see in this figure.

\begin{figure}
    \dofig{5.in}{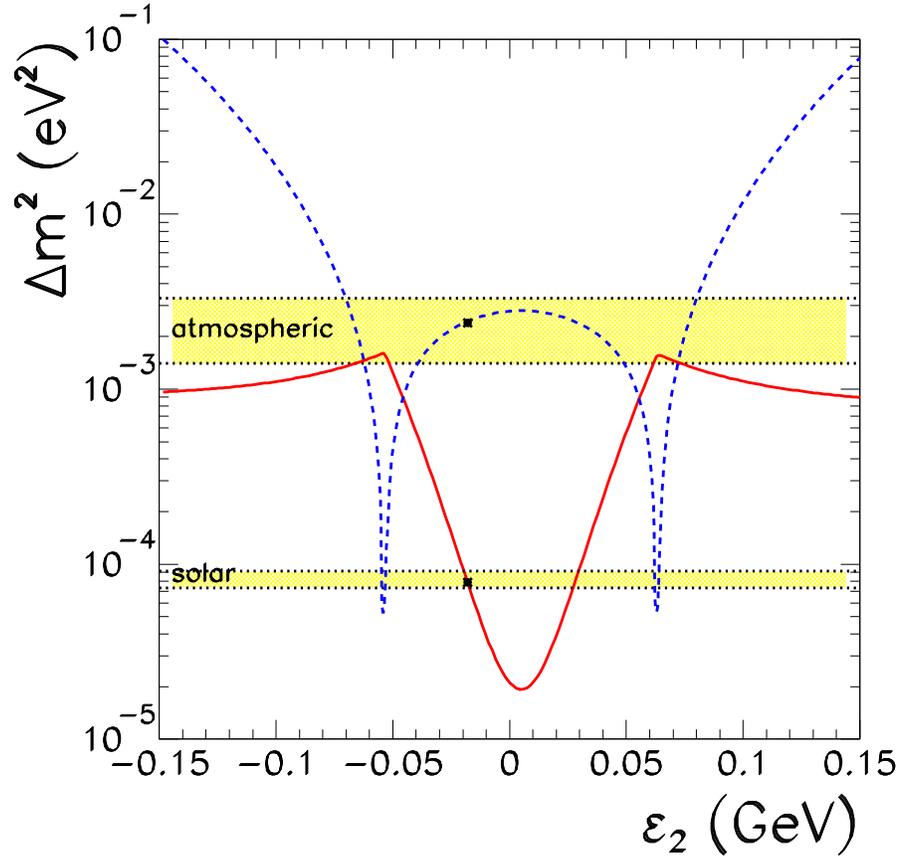}
\caption[]{
  The red solid (blue dashed) line stands for the atmospheric (solar) mass
  squared difference as a function of $\epsilon_2$ with the remaining
  parameters fixed at their reference value.
 The allowed $3\sigma$ atmospheric (solar) mass
  squared difference is represented by the upper (lower) horizontal yellow
  band.  Our reference point is represented by a star. }
\label{e2:dm}
\end{figure}

Fig.~\ref{e2:dm} presents the dependence on $\epsilon_2$ of the predicted
atmospheric neutrino squared mass difference $\Delta m^2_{\mathrm{atm}}$ (red
solid line) and the solar neutrino squared mass difference $\Delta
m^2_{\mathrm{sol}}$ (blue dashed line) for all other parameters fixed at the
reference values.  The cusps we observe in this figure are due to the crossing
of eigenvalues of the effective neutrino mass matrix and the fact that we
order the neutrino masses $m_{\nu_1} < m_{\nu_2} < m_{\nu_3} $.  Moreover, it
is clear from this figure that the solar neutrino squared mass difference
excludes a larger fraction of the parameter space than the corresponding
atmospheric difference; not only the solar squared mass difference is more
precisely known, but also the radiative corrections exhibit a strong
dependence on $\epsilon_2$. Nevertheless, there are still two regions where
both constraints are satisfied simultaneously. In the region of this figure
where the values of $|\epsilon_2|$ are small we can understand the observed
dependence on $\epsilon_2$: The key element is the fact that the term
dependent on $B$ in the square root in Eq.~(\ref{numasses}) is numerically
smaller than the term dependent on $A$. When this condition is fulfilled the
following approximated expressions for the two heaviest neutrinos are valid
for $\Lambda_2=\Lambda_3$
\begin{equation}
  m_{\nu_{2,3}}\simeq
  A\Lambda_3^2+B\Lambda_3(\epsilon_3+\epsilon_2)\pm A\Lambda_3^2\,,
\end{equation}
and as before, we approximate the atmospheric and the solar mass differences
by $m_{\nu_3}^2$ and $m_{\nu_2}^2$ respectively. In this way, the quadratic
growing of the solar mass with $\epsilon_2$ is clear, and the quadratic
decreasing of the atmospheric mass is also clear knowing that the term
proportional to $A$ is positive and larger than the negative term proportional
to $B$.

\begin{figure}
    \dofig{5.in}{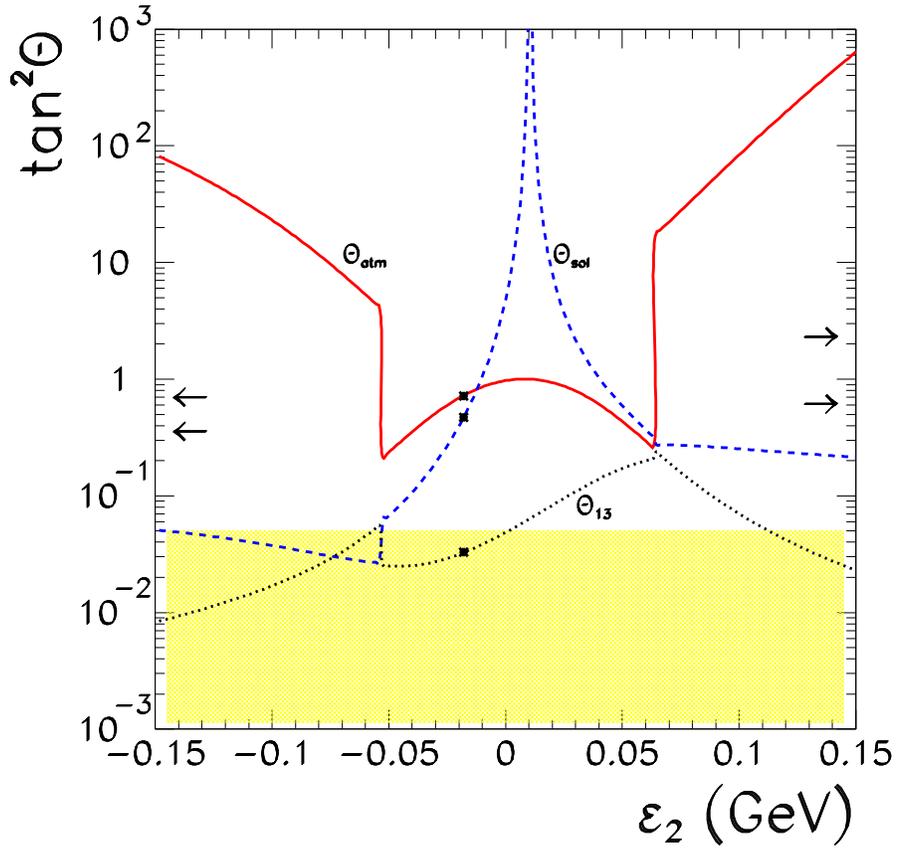}
\caption[]{
  The solid (dashed, dotted) line stands for the $\tan^2 \theta_{\mathrm{ atm
      (solar, 13)}}$ as a function of $\epsilon_2$ with the remaining
  parameters fixed at their reference value. The arrows on the right (left) of
  the figure indicate the $3\sigma$ bounds for atmospheric (solar) neutrinos.
  The CHOOZ allowed region for $\tan\theta_{13}$ is represented by the
  horizontal yellow area.  Our reference point is marked by a star.}
\label{e2:tan}
\end{figure}

We display in Fig.~\ref{e2:tan} the mixings $\tan^2 \theta_{\mathrm{atm (sol,
    13)}}$ as a function of $\epsilon_2$ for our reference point.  Once again,
we can see clearly the crossing of the eigenvalues of the effective neutrino
mass matrix that leads to the cusps in this figure. Furthermore, we observe a
sharp peak on the solar angle at small values of $\epsilon_2$ which is well
explained by Eq.~(\ref{mix:solar}) that predicts that $\tan
\theta_{\mathrm{sol}}$ diverges for $\epsilon_2 = \epsilon_3
\Lambda_2/\Lambda_3$, {\em i.e.}  $\epsilon_2 = \epsilon_3 = 0.011$ GeV for
our reference point.  The one--loop approximation (\ref{mix:solar}) also
describes well the solar mixing for our reference point predicting that
$\tan^2 \theta_{\mathrm{sol}} \simeq 0.43$. Conversely, the behavior of $\tan
\theta_{\mathrm{atm}}$ with $\epsilon_2$ is well described by the approximated
expression (\ref{tanatmapp2}) which predicts $\tan^2 \theta_{\mathrm{atm}}=
0.68$ (1) for our reference point ($\epsilon_2=\epsilon_3$). Lastly, radiative
corrections are also important for $\tan^2 \theta_{13}$ to satisfy the CHOOZ
bounds since, for our reference point, the tree--level prediction is $\tan^2
\theta_{13}\simeq 0.06$; too close to the CHOOZ bound.

\begin{figure}
    \dofig{5.in}{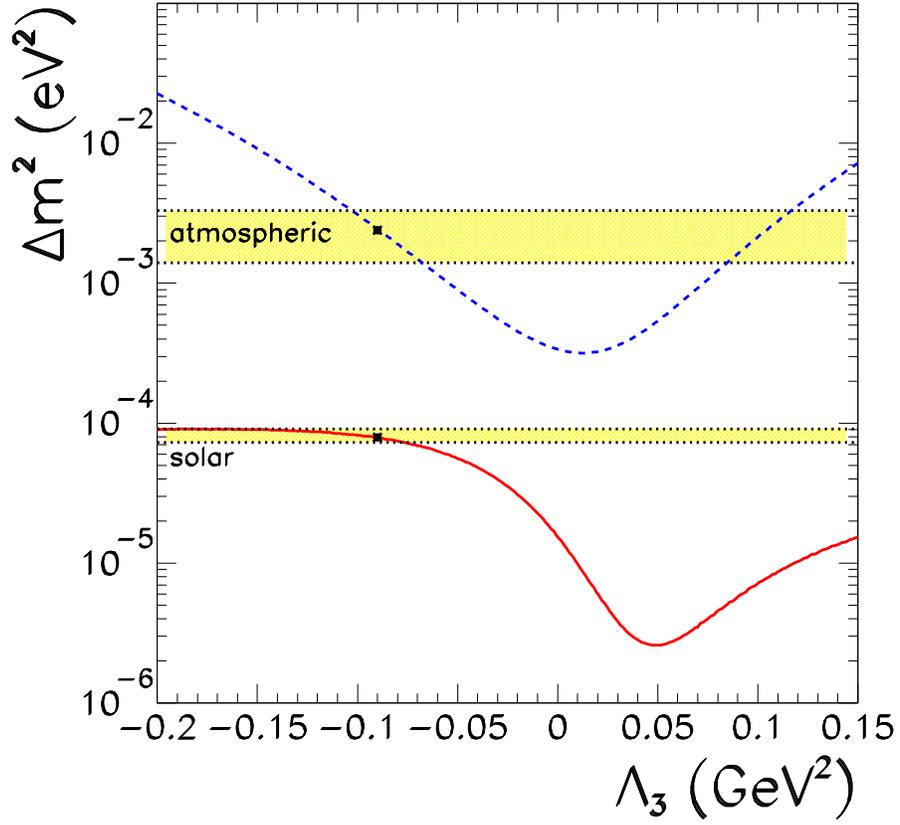}
\caption[]{
  The solid (dashed) line stands for the atmospheric (solar) mass
  squared difference as a function of $\Lambda_3$ with the remaining
  parameters fixed at their reference value. The upper (lower) shaded
  area corresponds to the $3 \sigma$ allowed region for the
  atmospheric (solar) mass squared difference.  Our reference point is
  represented by a star. }
\label{L3:dm}
\end{figure}

Fig.~\ref{L3:dm} contains the neutrino mass square differences as a function
of $\Lambda_3$ for all other parameters fixed at the reference value; the
behavior of $\Delta m^2$ is similar for $\Lambda_1$ and $\Lambda_2$. In this
case there is no neutrino mass eigenvalue crossing which reflects in the
absence of cusps in this figure. Moreover, we can see that the atmospheric
mass squared difference is much more sensitive to $\Lambda_3$ since it
receives a large tree--level contribution.

\begin{figure}
    \dofig{5.in}{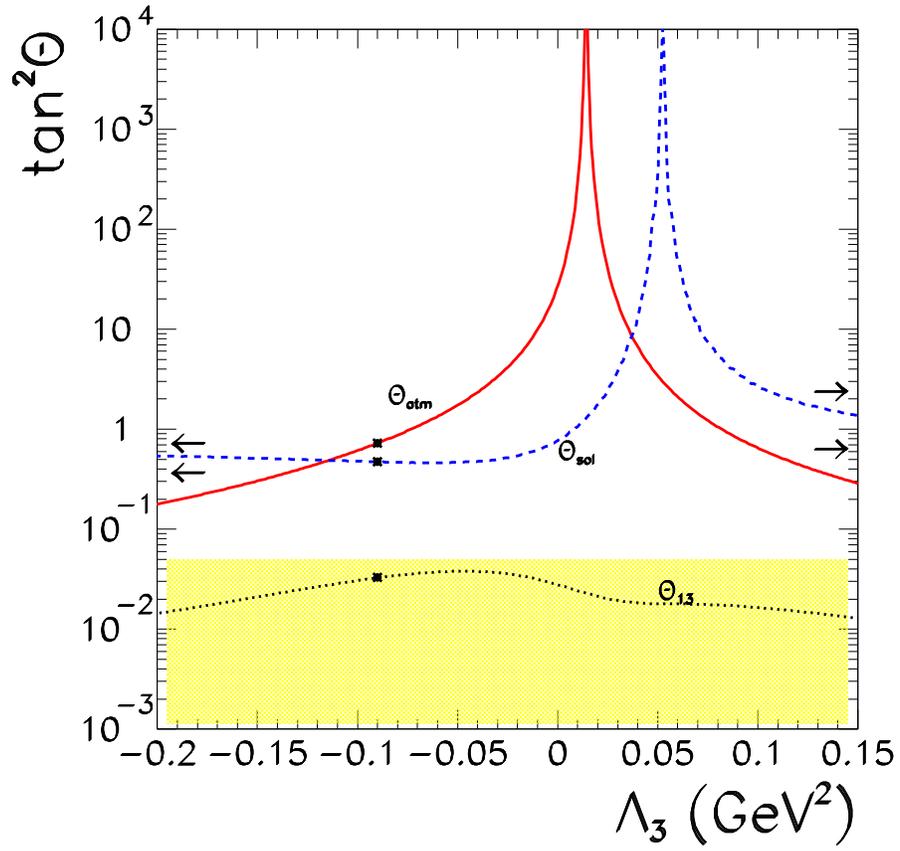}
\caption[]{ Mixing angles as a function of $\Lambda_3$. The conventions are 
  the same of Fig.~\ref{e2:tan}.  }
\label{L3:tan}
\end{figure}

We present in Fig.~\ref{L3:tan} the mixing angles as a function of $\Lambda_3$
for our reference point. Once again, the sharp peak in $\tan^2
\theta_{\mathrm{sol}}$ is well explained by Eq.~(\ref{mix:solar}) that
predicts the peak at $\Lambda_3 = \Lambda_2 \epsilon_3/\epsilon_2$, {\em i.e.}
$\Lambda_3 = 0.055$ GeV$^2$ for our reference point.  Another interesting
feature of this Figure is the peak in $\tan \theta_{\mathrm{atm}}$ which is
not predicted neither by the tree--level approximation (\ref{mix:atm}) nor the
approximated expression taking into account one--loop effects
(\ref{tanatmapp}).  This can be understood as follows. In the approximated
formula Eq.~(\ref{tanatmapp}) we neglected terms that are small for our
reference point. If $|\Lambda_3|$ is decreased, some of the neglected terms in
the 3,2 entry of ${\bf M}^{\mathrm{eff}}$, see Eq.~(\ref{massmat}), are no
longer negligible. A better approximation for the atmospheric angle is, in
this case,
\begin{equation}
  \tan2\theta_{\mathrm{atm}}=\frac
  {2A\Lambda_2\Lambda_3+2B(\epsilon_3\Lambda_2+\epsilon_2\Lambda_3)}
  {A(\Lambda_3^2-\Lambda_2^2)+2B(\epsilon_3\Lambda_3-\epsilon_2\Lambda_2)}
  \; .
\label{tanatmapp3}
\end{equation}
which predicts a peak for $\tan\theta_{\mathrm{atm}}$ at
\begin{equation}
\Lambda_3\approx\frac{-B\epsilon_3\Lambda_2}
{A\Lambda_2+B\epsilon_2}\sim 0.01 \,{\mathrm{GeV}}^2
\end{equation}
which agrees very well with the numerical result.  Clearly the peaks in this
figure signals the necessity of using the full neutrino effective mass matrix,
as well as the importance of the radiative corrections.

\section{Discussion}
\label{bye}

In this previous section we showed that an AMSB model with bilinear
$R$--parity violation is viable, {\em i.e.} we exhibit a region of the
parameter space where the model predicts neutrino masses and mixings in
agreement with the experimental data as well as the superpartner masses are
heavier than the available experimental constraints.  Certainly, the most
challenging point for this class of models is to generate a solar neutrino
mass squared difference in agreement with data since $\Delta m^2_{\rm solar}$
is due to radiative corrections that depend strongly on the superpartners
masses. On the other hand, the connection between the SUSY spectrum,
$R$--parity violating parameters and neutrino properties allow us to directly
probe this class of models in collider experiments because the $R$--parity
violating interactions not only might lead to a decay inside the detector of
the lightest SUSY particle (LSP), but also it can originate new decay channels
for the other superpartners. Furthermore, more precise determinations of the
neutrino masses and mixing angles can be also used to falsify this kind of
models due to sensitivity of the solar parameters to radiative corrections,
and hence to the model parameters. For instance, our reference point can be
ruled out by improved limits, or an actual measurement, of $\tan \theta_{13}$.

In the AMSB--BRpV framework, the lightest neutralino presents leptonic decays
$ \tilde{\chi}^0_1 \to \nu \ell^+ \ell^{\prime-}$, semi-leptonic ones $
\tilde{\chi}^0_1 \to \nu q \bar q $ or $ \ell q \bar q$, and the invisible
mode $\tilde{\chi}^0_1 \to \nu \nu \nu$. If its decay occurs inside the
detector we have to take into account new topologies in the search for SUSY
since the missing transverse energy is reduced in this scenario as well as
there is a larger production of leptons and jets \cite{Magro:2003zb}. In
addition to that, the LSP can give rise to displaced vertices which can be
used as a smoking gun for SUSY \cite{porod}; see \cite{Hirsch:2003fe} for more
studies relating neutrino and collider physics.  Moreover, in the case of AMSB
the lightest neutralino and chargino are almost degenerate, implying that the
$R$--parity violating interactions may play an important role in the decay of
the lightest chargino ($\tilde{\chi}_1^\pm$). The possible new
$\tilde{\chi}_1^\pm$ decays induced by BRpV are
\begin{eqnarray}
   \tilde{\chi}_1^\pm &\to& \bar{q} q^\prime \nu_i \; ,
\nonumber \\
   \tilde{\chi}_1^\pm &\to& \ell^\pm q \bar{q}  \; ,
\nonumber \\
   \tilde{\chi}_1^\pm &\to& \ell^+_i \ell^-_j \ell^\pm_k \; ,
 \\
   \tilde{\chi}_1^\pm &\to& \ell^\pm_i \nu_j \nu_k   \; ,
\nonumber
\end{eqnarray}
where $\ell_i = e,~\mu,~\tau$. In order to access the importance of
these new signatures for AMSB further detailed studies are needed \cite{prep}.


\acknowledgments

This work was supported in part by {\bf Conicyt grant No.~1040384}, by
Conselho Nacional de Desenvolvimento Cient\'{\i}fico e Tecnol\'ogico
(CNPq), and by Funda\c{c}\~ao de Amparo \`a Pesquisa do Estado de
S\~ao Paulo (FAPESP). 

%


%

%


\end{document}